# The OPFOS microscopy family: High-resolution optical-sectioning of biomedical specimens

Jan A.N. Buytaert[1,°], Emilie Descamps[2], Dominique Adriaens[2], and Joris J.J. Dirckx[1]

[1] Laboratory of BioMedical Physics – University of Antwerp, Groenenborgerlaan 171, B-2020 Antwerp, Belgium
[2] Evolutionary Morphology of Vertebrates – Ghent University, K.L. Ledeganckstraat 35, B-9000 Gent, Belgium

° Corresponding author:
email jan.buytaert@ua.ac.be;
telephone 0032 3 265 3553;
fax 0032 3 265 3318.

# Abstract

We report on the recently emerging (Laser) Light Sheet based Fluorescence Microscopy field (LSFM). The techniques used in this field allow to study and visualize biomedical objects non-destructively in high-resolution through virtual optical sectioning with sheets of laser light. Fluorescence originating in the cross section of the sheet and sample is recorded orthogonally with a camera.

In this paper, the first implementation of LSFM to image biomedical tissue in three dimensions – Orthogonal-Plane Fluorescence Optical Sectioning microscopy (OPFOS) – is discussed. Since then many similar and derived methods have surfaced (SPIM, Ultramicroscopy, HR-OPFOS, mSPIM, DSLM, TSLIM, …) which we all briefly discuss. All these optical sectioning methods create images showing histological detail.

We illustrate the applicability of LSFM on several specimen types with application in biomedical and life sciences.

# Introduction

Serial (Mechanical) Histological Sectioning (SHS) creates physical slices of fixed, stained and embedded tissues which are then imaged with an optical microscope in unsurpassed sub-micrometer resolution. Obtaining these slices is however extremely work-intensive, requires physical (one-time and one-directional) slicing and thus destruction of the specimen. A 2-D sectional image reveals lots of histologically relevant information, but a data stack and its 3-D reconstruction are even more essential for the morphological interpretation of complex structures, because they give additional insight in the anatomy. The SHS method requires semi-automatic to manual image registration to align all recorded 2-D slices in order to get realistic 3-D reconstructions. Often dedicated image processing of the sections is needed because of the geometrical distortions from the slicing.

A valuable alternative to achieve sectional imaging and three-dimensional modeling of anatomic structures can be found in the little known and relatively recent field of microscopy called (Laser) Light Sheet based Fluorescence Microscopy or LSFM. These non-destructive methods generate registered optical sections in real-time through bio(medical) samples ranging from microscopic till macroscopic size. LSFM can reveal both bone and soft tissue at a micrometer resolution, thus showing a large amount of histological detail as well.

The first account of the LSFM idea was published by Voie, Burns and Spelman in 1993 and applied to image the inner ear cochlea of guinea pig [1]. Their method was called Orthogonal-Plane Fluorescence Optical Sectioning (OPFOS) microscopy or tomography. The motivations for the OPFOS invention were (1) the above mentioned disadvantages of serial histological sectioning, (2) the typical photo-bleaching of fluorophores in conventional or confocal fluorescence microscopy, and (3) the fact that samples are optically opaque which means a limited penetration depth and inefficient delivering and collecting of light.

Surprisingly, all these problems can be avoided by combining two old techniques. Voie and colleagues first combined the Spalteholz method of 1911 [2] with the even older Ultramicroscope method of 1903 [3]. In most microscopy techniques, the same optical path and components are used for the illumination and the observation of light. Siedentopf and Nobel Prize winner Zsigmondy made a simple change of the optical arrangement in their Ultramicroscopy setup by separating the illumination and viewing axis [3]. Furthermore, their illumination was performed by a thin plane or sheet of light. Orthogonal viewing or observation of this sheet offers full-field and real-time sectional information. Their method was originally developed for gold particle analysis in colloidal solutions with sunlight. OPFOS used the same optical arrangement but for tissue microscopy. The separation of the illumination and imaging axis combined with laser light sheet illumination only illuminates the plane that is under observation (in contrast to confocal microscopy) and thus avoids bleaching in sample regions that are not being imaged. Generally, samples are optically opaque so the plane of laser light cannot section the sample. Spalteholz introduced a clearing method which dates back exactly 100 years [2]. His museum technique is capable of making tissue transparent by matching the refractive index throughout the entire object volume by means of a mixture of oils with refractive indices close to that of protein. Submerged in this Spalteholz fluid, a prepared specimen appears invisible, with light passing right through it unscattered and without absorption. This clearing or refractive index matching is essential for the OPFOS technique to achieve a penetration depth of several millimeters. This procedure is followed by staining of the sample with fluorescent dye or just by just relying on naturally occurring auto-fluorescence. The sectioning laser plane activates the fluorophores in the cross section of sheet and sample, which are finally orthogonally recorded by a camera.

OPFOS utilizes yet a third method in conjunction with the two previous techniques when the specimen contains calcified tissue or bone. In this case, the calcium first needs to be removed before the Spalteholz procedure is applied. Bone cannot be made transparent, as the calcium atoms strongly scatter light.

Since 1993, many OPFOS-like derived methods were developed for tissue microscopy, all based on light sheet illumination. 'LSFM' has become a broadly accepted acronym to cover the whole of these techniques, coined in Dresden (Germany) 2009. In the discussion, we will give a short overview of this OPFOS-derived LSFM microscopy family. First, we will explain in detail the specimen preparation and the optical arrangement of the original OPFOS setup. The remainder of this paper will serve to demonstrate some applications of OPFOS.

# Materials and Methods

## Specimen preparation

In most LSFM methods, the biomedical tissue samples are severely limited in size, though for instance the LSFM implementations of Ultramicroscopy, HR-OPFOS and TSLIM (cf. the discussion section) are capable of imaging macroscopic samples up to tens of millimeters [4]. In all cases, an elaborate specimen preparation is required:

- Euthanasia: Living animals cannot be used in combination with clearing solutions. In general, LSFM is thus mainly used *in vitro.* Clearing can be omitted and living animals can be used if the species possesses a natural transparency at a certain developmental stage, for instance fish embryos [5,6]. The embryos are immobilized by embedding in agarose.
- Perfusion: Before dissecting a sample to the required dimensions, transcardial perfusion with phosphate buffered saline is useful as coagulated blood is difficult to clear with Spalteholz fluid [7-9]. If perfusion is omitted, bleaching is required.
- Fixation: Immersion in 4% paraformaldehyde (10% formalin) during 24h for preservation and fixation of the specimen.
- Bleaching: Optional bleaching in 5% to 10% hydrogen peroxide for one or more days can be performed when the sample contains dark pigmented tissue (e.g. black skin, fish eyes) [10]. This step can also be applied after decalcification [11].
- Decalcification: When the specimen contains calcified or mineralized tissue, such as cartilage or bone, decalcification is in order. A 10% demineralized water solution of dihydrate ethylenediaminetetraacetic acid (EDTA) slowly diffuses calcium atoms from the sample through a chelation process. Low power microwave exposure (without heating) drastically accelerates the decalcification process from a month to several days [12,13].

- Dehydration: Immersion in a graded ethanol series (f.i. 25%, 50%, 75%, 100%, 100% each for 24h) removes all water content from the sample [12]. In the final 100% step, optional addition of anhydrous copper sulfate at the bottom of the ethanol bath might improve the dehydration [14].
- Hexane or benzene: The optional immersion in a graded series of hexane or benzene is said to improve dehydration further [8,11,14,15]. Furthermore, hexane might assist in clearing myelin present in the tissue sample. Nerve axons are surrounded by myelin sheets which do not easily become transparent with Spalteholz fluid.
- Clearing: To achieve large volume imaging in inherently less transparent samples, clearing is needed. The specimen are to be immersed in clearing solution, either through a graded series (f.i. 25%, 50%, 75%, 100%, 100% each for 24h) when the hexane or benzene step was skipped [12], or directly in 100% pure clearing solution when hexane or benzene were applied [8,11]. The clearing solution mimics the refraction index of protein and matches the refraction index of the sample to the solution. The solution can either consist of pure benzyl benzoate followed in a later stage by the final mixture [14], or directly of this mixture solution. A 5:3 mixture of methyl salicylate and benzyl benzoate is called Spalteholz fluid [1,2,7]. For brain tissue, a 1:2 mixture of benzyl alcohol and benzyl benzoate has been found to give better results [8,11].
- Staining: The required fluorescence can originate from auto-fluorescence from lipofuscins, elastin and/or collagen [8]. Fluorescent staining can be applied by immersion in a dye bath (of f.i. Rhodamine B isothiocyanate in clearing solution [1]) or even by functional staining. However, many fluorescent dyes deteriorate or even break down completely because of the aggressive clearing solution used, e.g. GFP.

## Optical setup

In what follows, the original OPFOS setup is discussed as it was introduced by Voie et al. in 1993 [1]. Many improved versions have been developed since, all based on the OPFOS or Ultramicroscopy design (cf. the discussion).

The setup is represented in Figure 1 and Figure 2. The prepared sample is illuminated by an XY-sheet of laser light travelling along the X-axis. The omni-directional fluorescence light emitted in the positive Z-axis is used for imaging. Virtual section images in the XY-plane are hence recorded; by translation of the specimen along the Z-axis, an aligned sequence of section images is obtained.

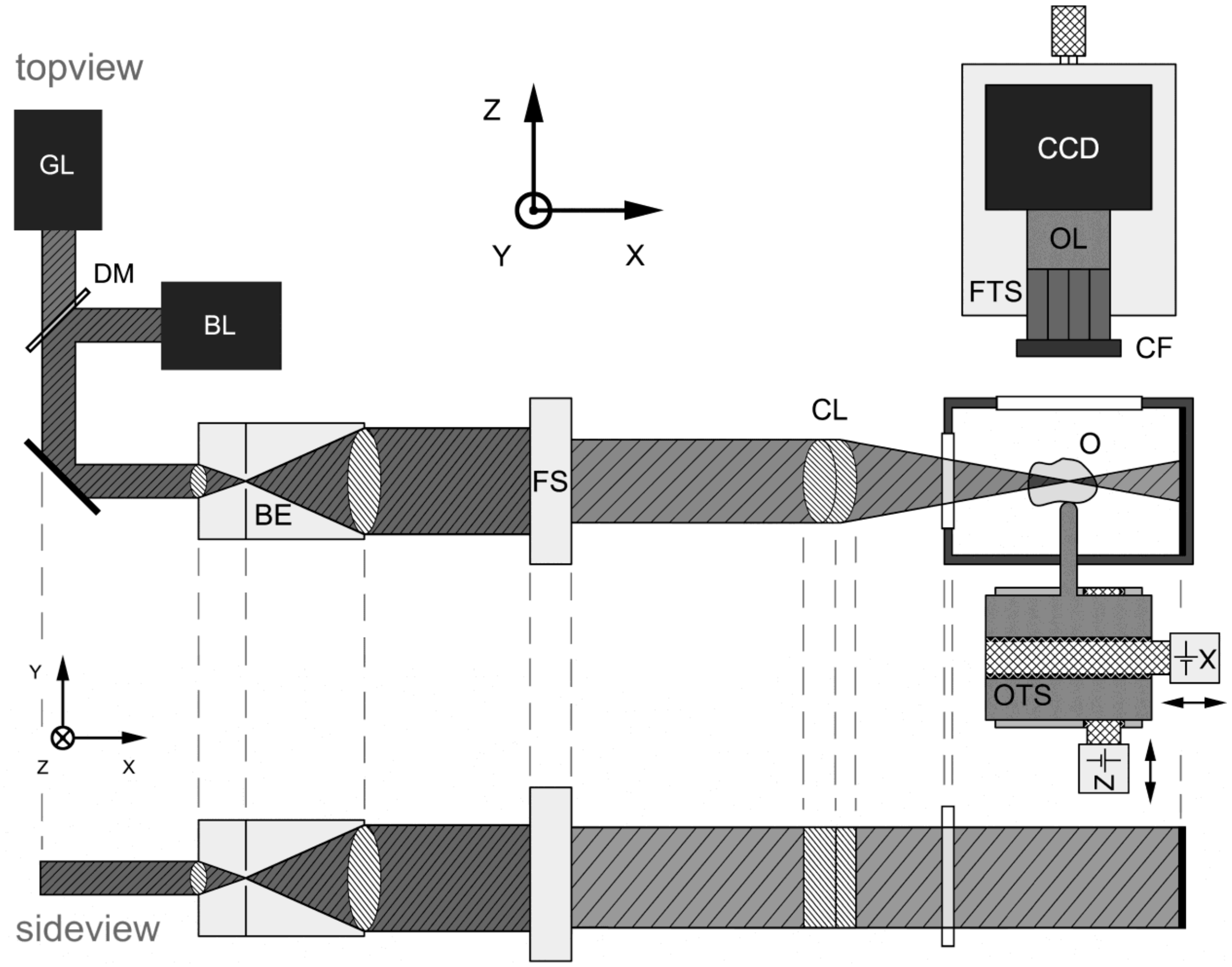

**Figure 1.** Schematic drawing of the (HR-)OPFOS setup: Light from a green (GL) or blue laser (BL) passes through a Keplerian beam expander (BE) with spatial filter, a field stop (FS) and a cylindrical achromat lens (CL) which focuses the laser along one dimension within the transparent and fluorescent object (O). A two-axis motorized object translation stage (OTS) allows scanning of the specimen and imaging of different depths. The fluorescence light emitted by the object, is projected onto a CCD-camera by a microscope objective lens (OL) with fluorescence color filter (CF) in front. The focusing translation stage (FTS) is used to make the objective lens focal plane coincide with the laser focus.

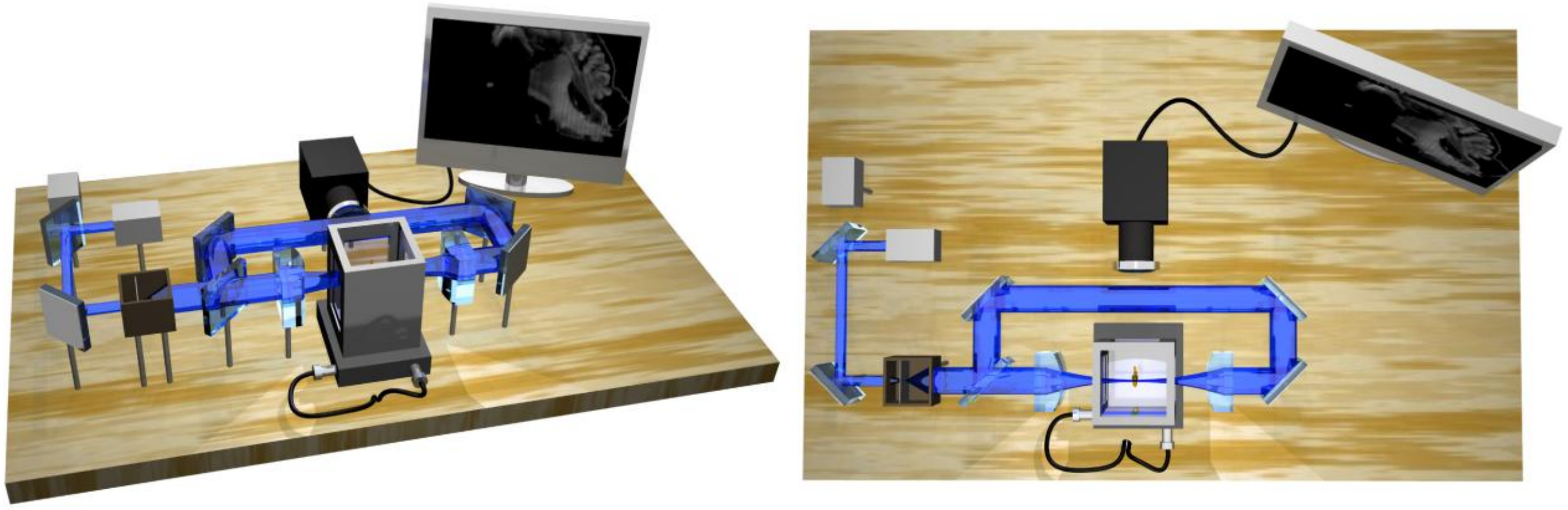

**Figure 2.** A 3-D setup representation of an (HR-)OPFOS setup with two-sided cylindrical lens sheet illumination, and with two laser wavelengths (green and blue). The blue laser is active here.

An essential requirement for OPFOS is the generation of a laser light sheet. In practice, it is impossible to generate a perfect plane or sheet of light; however, using a cylindrical lens a sheet can be approximated. A Gaussian laser beam is first expanded and collimated by a Keplerian beam expander. The broadened beam then travels along the X-axis through a cylindrical lens which focuses light in only one dimension to a line along the Z-axis. Along the Y-axis the Gaussian beam is unaltered, cf. Figure 1. In the XZ-plane, the light sheet has a hyperbolic profile in the focal zone, cf. Figure 3. The Z-thickness of the profile increases in either way along the X-axis when moving away from the minimal beam waist focus $d_1$. The Rayleigh range $x_R$ is the distance on either site of the minimal focus $d_1$ where the hyperbolically focused beam has thickened to $\sqrt{2}\,d_1$. This variable is described by the expression $b_1 = 2x_R = \frac{\pi d_1^2}{2\lambda}$, where $b_1$ is called the confocal parameter or the total distance in which a focus smaller than $\sqrt{2}d_1$ is maintained. The numerical aperture of the cylindrical lens is inversely related to the confocal parameter $b_1$ and directly proportional to the beam waist focal thickness $d_1$.

The height of the beam in Y-direction combined with the confocal parameter $b_1$ along the X-axis defines the size of the XY-sheet which sections the sample. The specimen consequently has to fit within this zone. A trade-off exists between maximal image and sample width ($\approx b_1$) and the sectioning thickness $\sqrt{2}\,d_1$ (~$1/\sqrt{b_1}$ ).

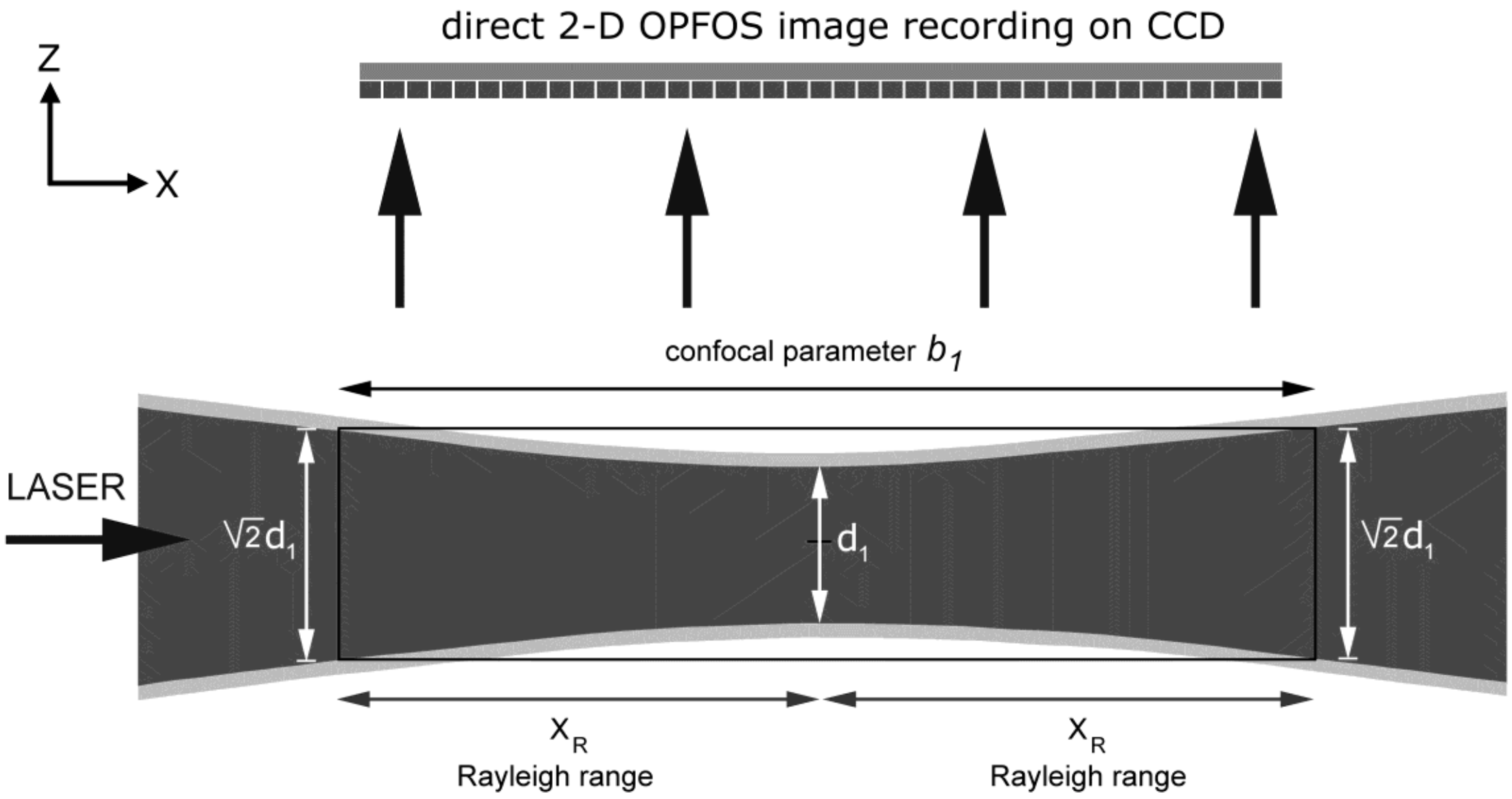

**Figure 3.** The hyperbolic focus profile of a cylindrical lens is shown. OPFOS records 2-D images in an approximated planar sheet defined by the confocal parameter zone $b_1$ where the thickness is considered constant at $\sqrt{2}d_1$. The dark gray area in the center represents the *$1/e^2$* intensity profile.

In summary, an OPFOS image has a slicing thickness $d_1$ in the center, growing to $\sqrt{2}\, d_1$ at the edges $x_R$. Everything within the thickness of the laser light sheet is integrated into a flat section image, so actually a varying thickness and slicing resolution is integrated in the 2-D image. The wavelength of the laser light depends on the fluorophore that is to be exited. A green laser (532 nm) is suited to excite Rhodamine B, while the blue laser (488 nm) is suited to evoke autofluorescence in many biomedical tissue samples.

# Results and Discussion

## Application examples

### Biomechanics of hearing

As a first illustration of the above described OPFOS setup, we show an application in hearing research of the middle ear [16]. Better understanding of the biomechanics of hearing through finite-element modeling requires accurate morphology of the hearing bones and their suspensory soft tissue structures. In Figure 4 and Figure 5, OPFOS cross sections in gerbil (*Meriones unguiculatus)* middle ears are shown, which can be segmented and triangulated into 3-D surface mesh models, cf. Figure 6. Thanks to the OPFOS technique, the sections through the sample can be visualized in real-time and clearly show histological detail on both bone and soft tissue.

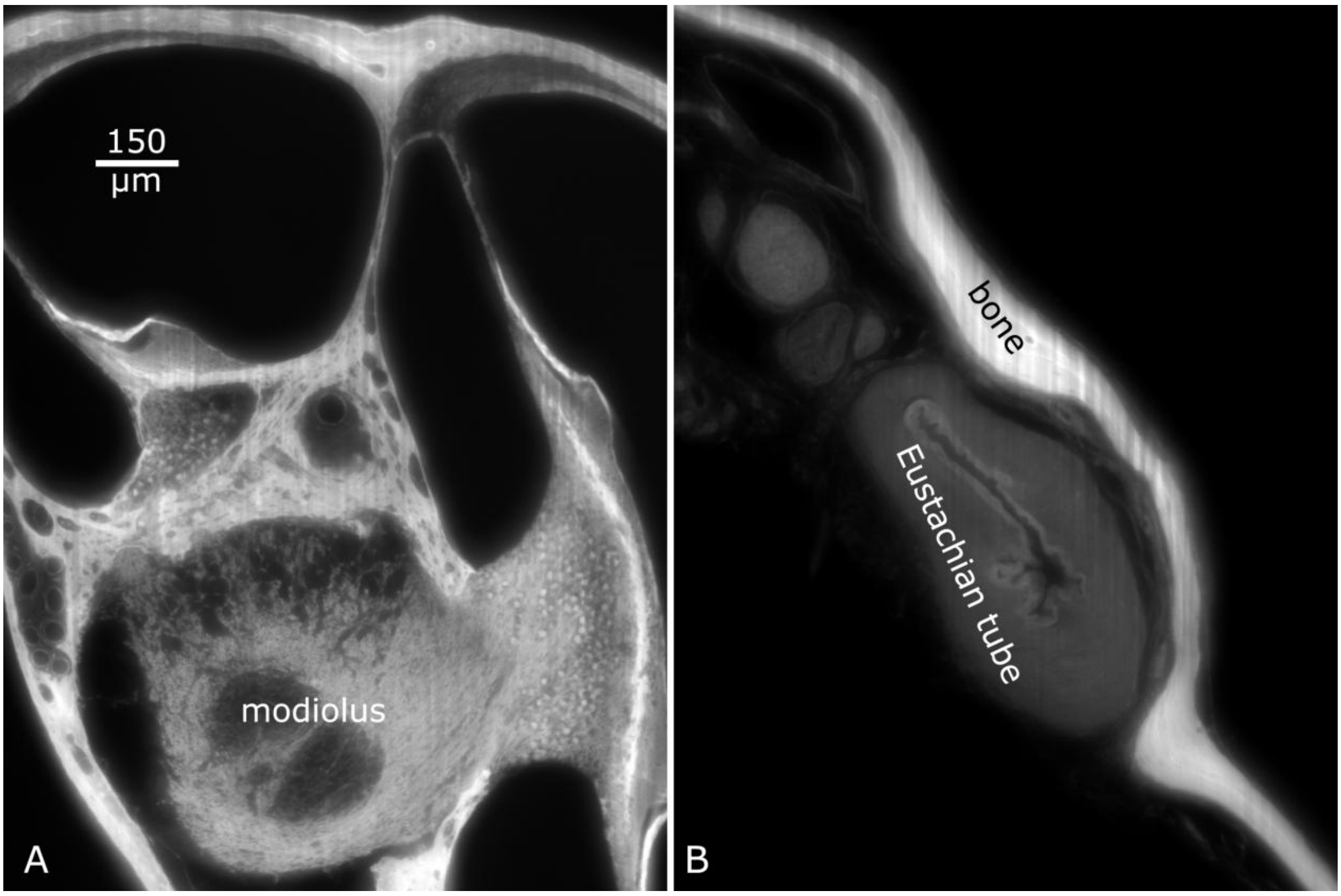

**Figure 4.** An OPFOS cross section of 1600x1200 pixels through **(A)** the scalae and modiolus of a gerbil inner ear cochlea and **(B)** a closed Eustachian tube in the middle ear. Rhodamine staining was combined with 532nm laser light sheet sectioning.

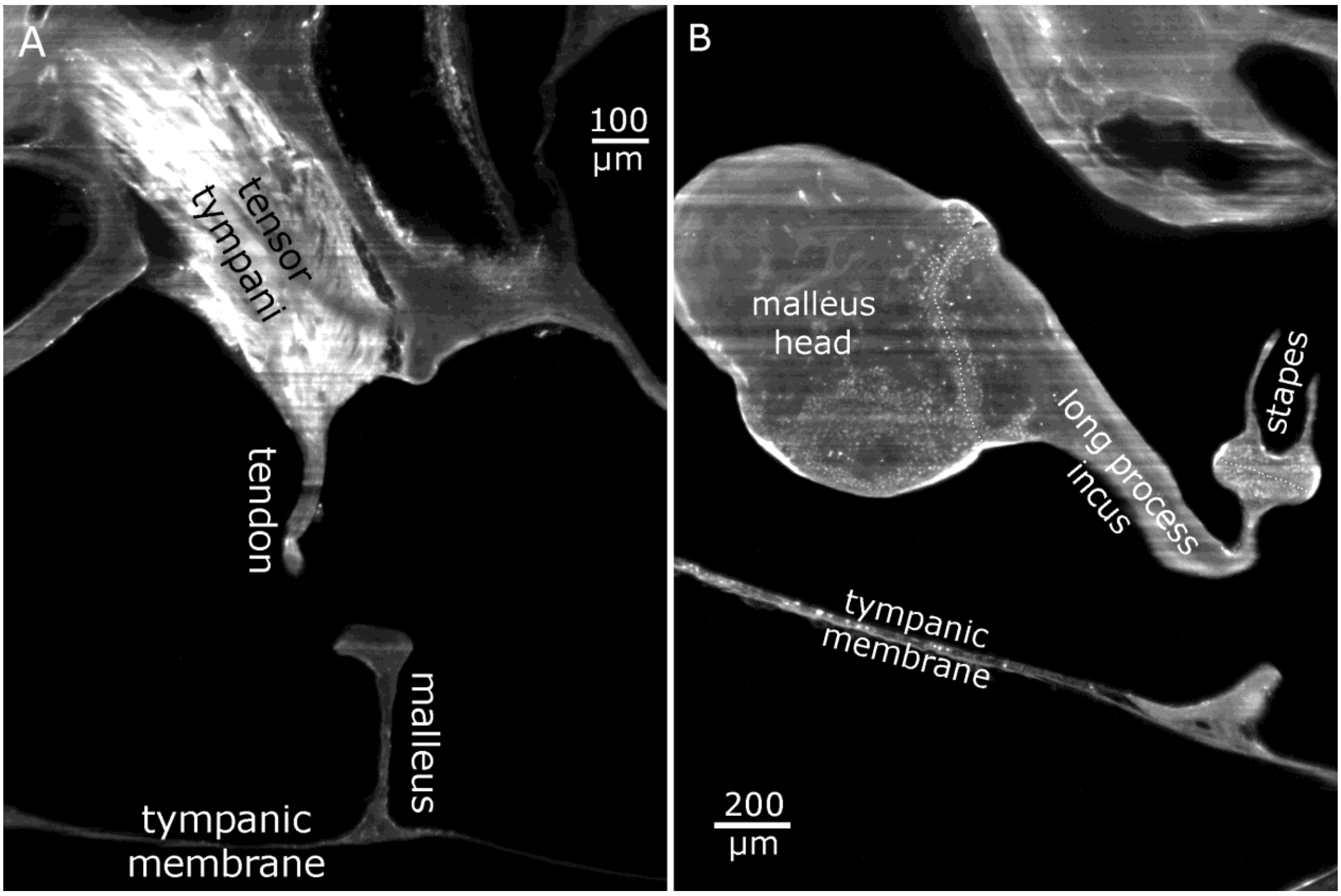

**Figure 5.** 2-D virtual cross sections (1600x1200 pixels) from OPFOS microscopy on the gerbil middle ear. **(A)** Tensor tympani muscle and tendon reaching down towards the malleus hearing bone. **(B)** Incudomallear and incudostapedial articulation between incus and malleus hearing bone. Rhodamine staining was combined with 532nm laser light sheet sectioning. Pixel size 1.5x1.5 µm.

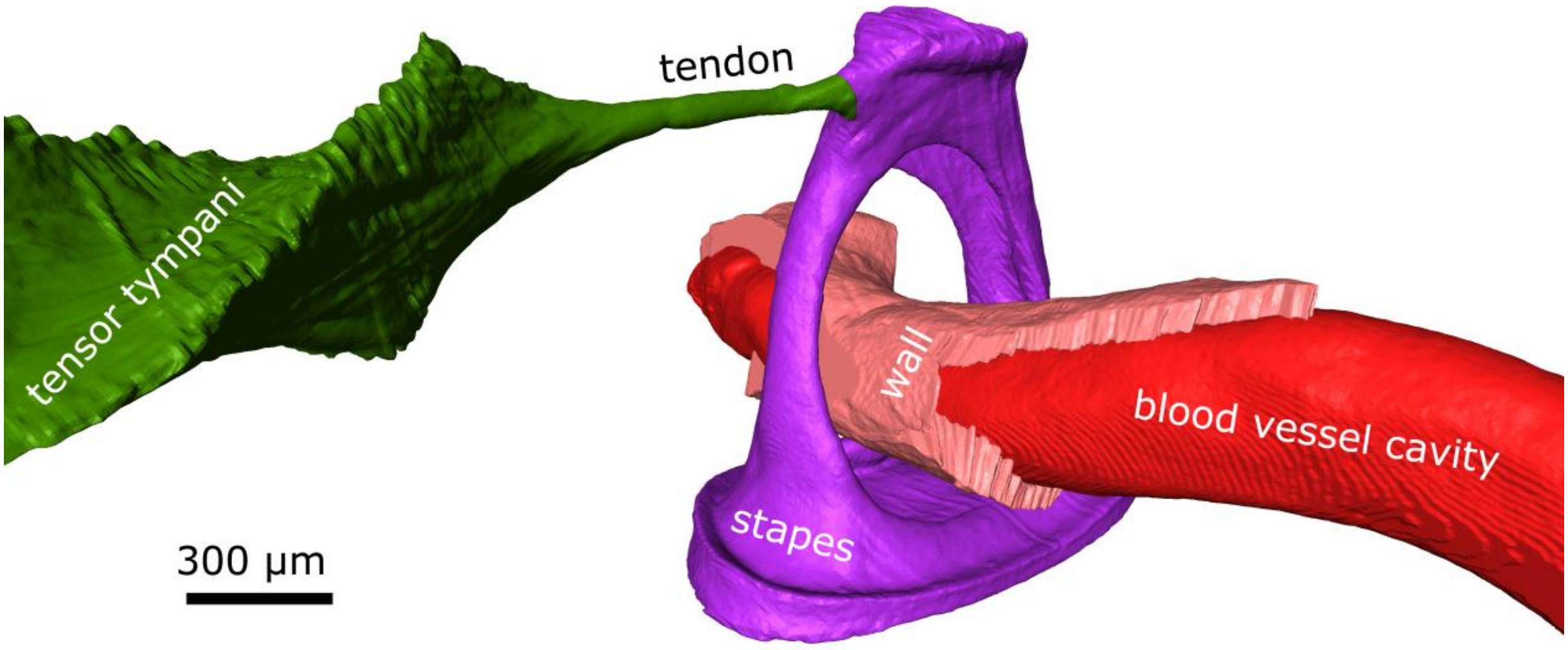

**Figure 6.** A 3-D OPFOS reconstruction showing a surface mesh of the stapes hearing bone, a blood vessel running through it, and the tensor tympani muscle attaching to the stapes head. The blood vessel wall and inner cavity are both separately modeled. Voxel size 1.5x1.5x5 µm.

**Morphology of the brain**

In neurology, morphological brain atlases are a useful tool. To this end, sectional imaging with histological detail of mice (C57 black *Mus musculus*) brain can be achieved with the OPFOS method, cf. Figure 7. The brain was cleared using the Spalteholz method, though for better results a combination of benzyl alcohol and benzyl benzoate could be used (cf. the section on specimen preparation). An extra hexane immersion step might further improve clearing of the brain.

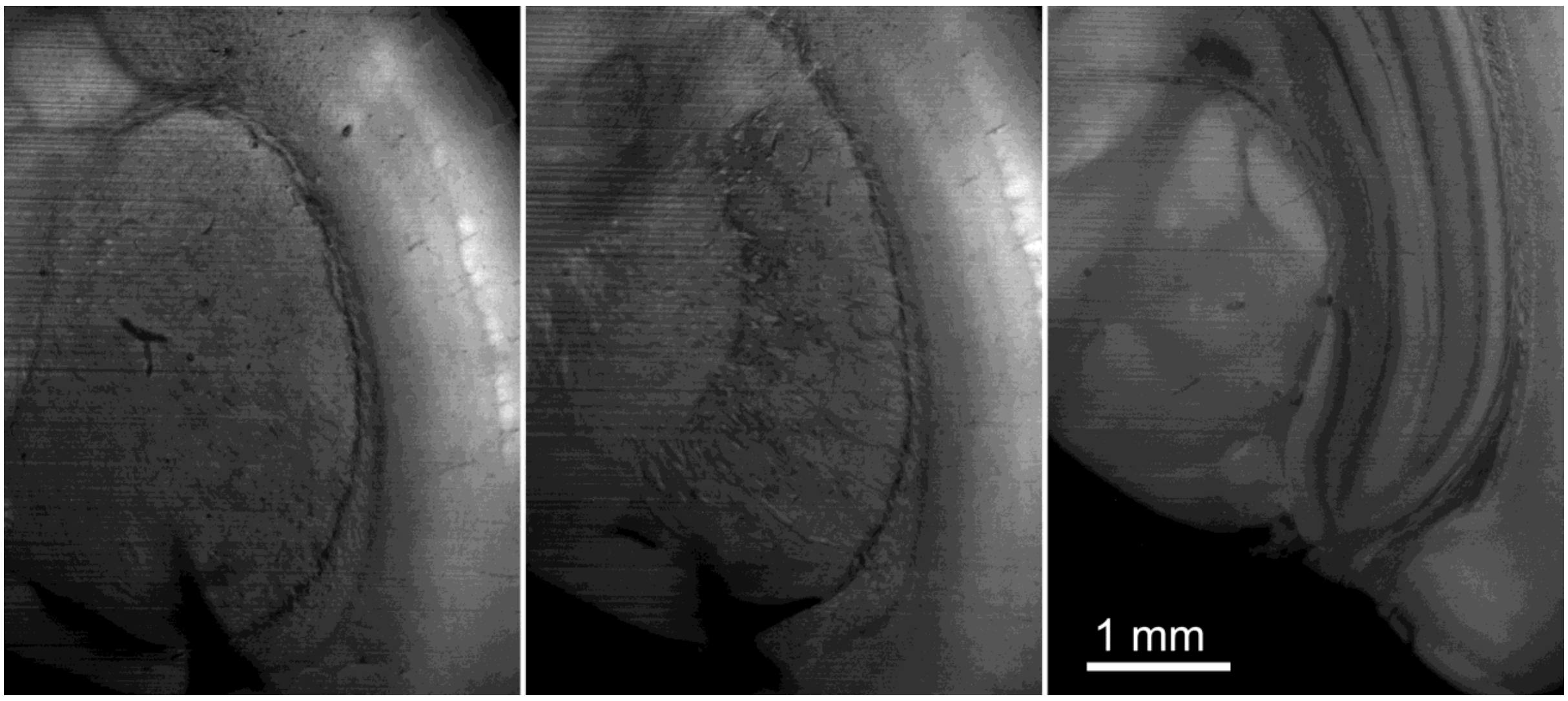

**Figure 7.** Three OPFOS cross sections of 1600x1200 pixels at different depths in a mouse brain. Natural autofluorescence of the brain was achieved using 488 nm laser light sheet sectioning. Pixel size 3x3 µm.

**Biomechanics of small vertebrates**

In morphological studies, functionality of a musculoskeletal system requires the visualization of both skeleton and muscles. For example, to gain insight in the feeding mechanisms of newly born seahorses (*Hippocampus reidi*), the shape, volume and orientation of the sternohyoideus muscle is, among other structures, of special interest (Figure 8). This conspicuous muscle spans from the shoulder girdle to the hyoid bar, assisting in an extremely rapid feeding strike in order to suck in prey [17].

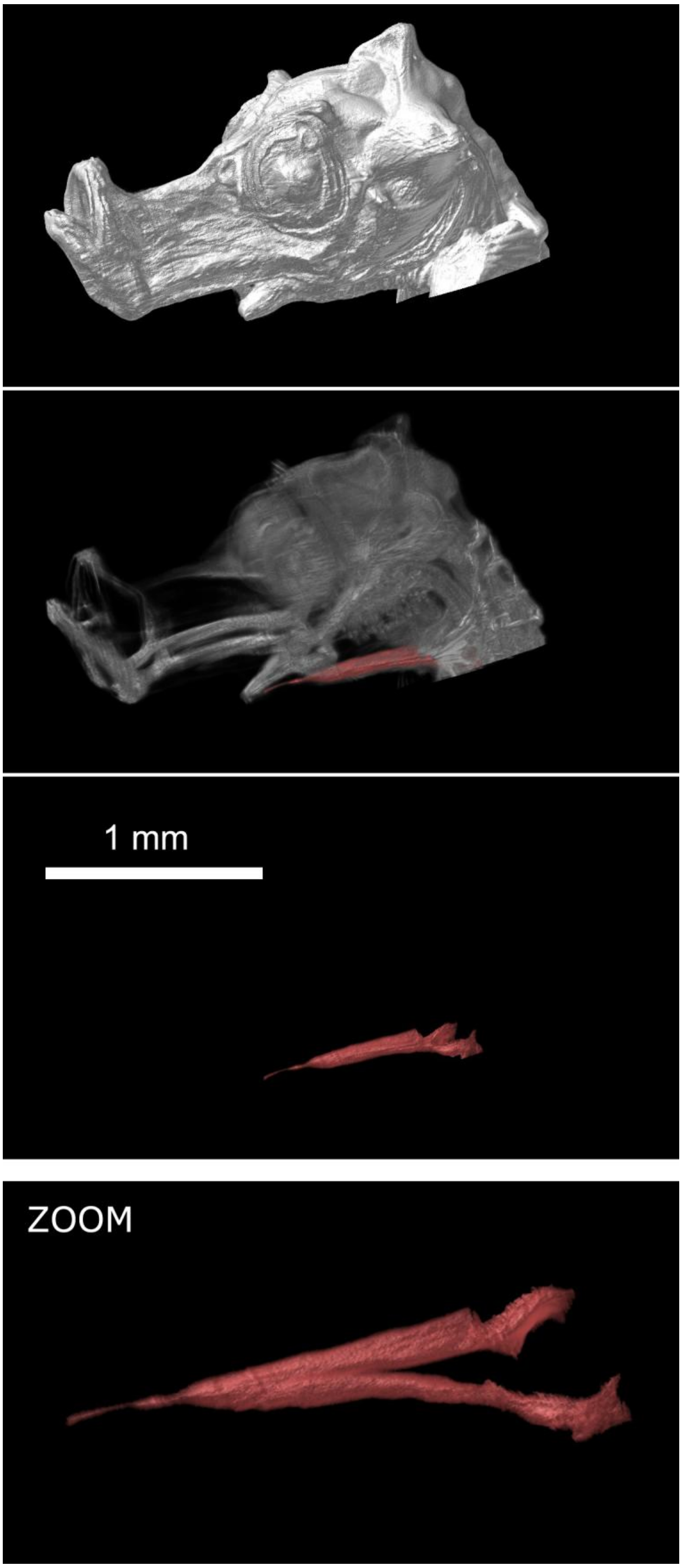

**Figure 8.** 3-D reconstruction of the head of a one-day old seahorse. The OPFOS image data is functionally segmented to study the morphology of the sternohyoideus muscle, cf. zoom (oblique view of the muscle). Natural autofluorescence of the head was achieved using 488 nm laser light sheet sectioning. Voxel size 3.5x3.5x5 µm.

Organogenesis and evolutionary morphology can benefit from OPFOS as well. The technique allows to discern the main structural elements of the head of an African clawed tadpole (*Xenopus laevis*) without any dissection. All different tissue types, such as muscle, skeletal and nervous tissues could be visualized and discriminated by their distinct (auto)fluorescence gray scales, cf. Figure 9. Skeletal structures were depicted as the darkest mass, corresponding to the lowest autofluorescence. By contrast, the nervous system (brain) was the brightest part, and to a lesser extent also the muscles showed high fluorescence. A 3-D reconstruction based on the gray scales in the OPFOS image stacks illustrates this in Figure 9 and Figure 10.

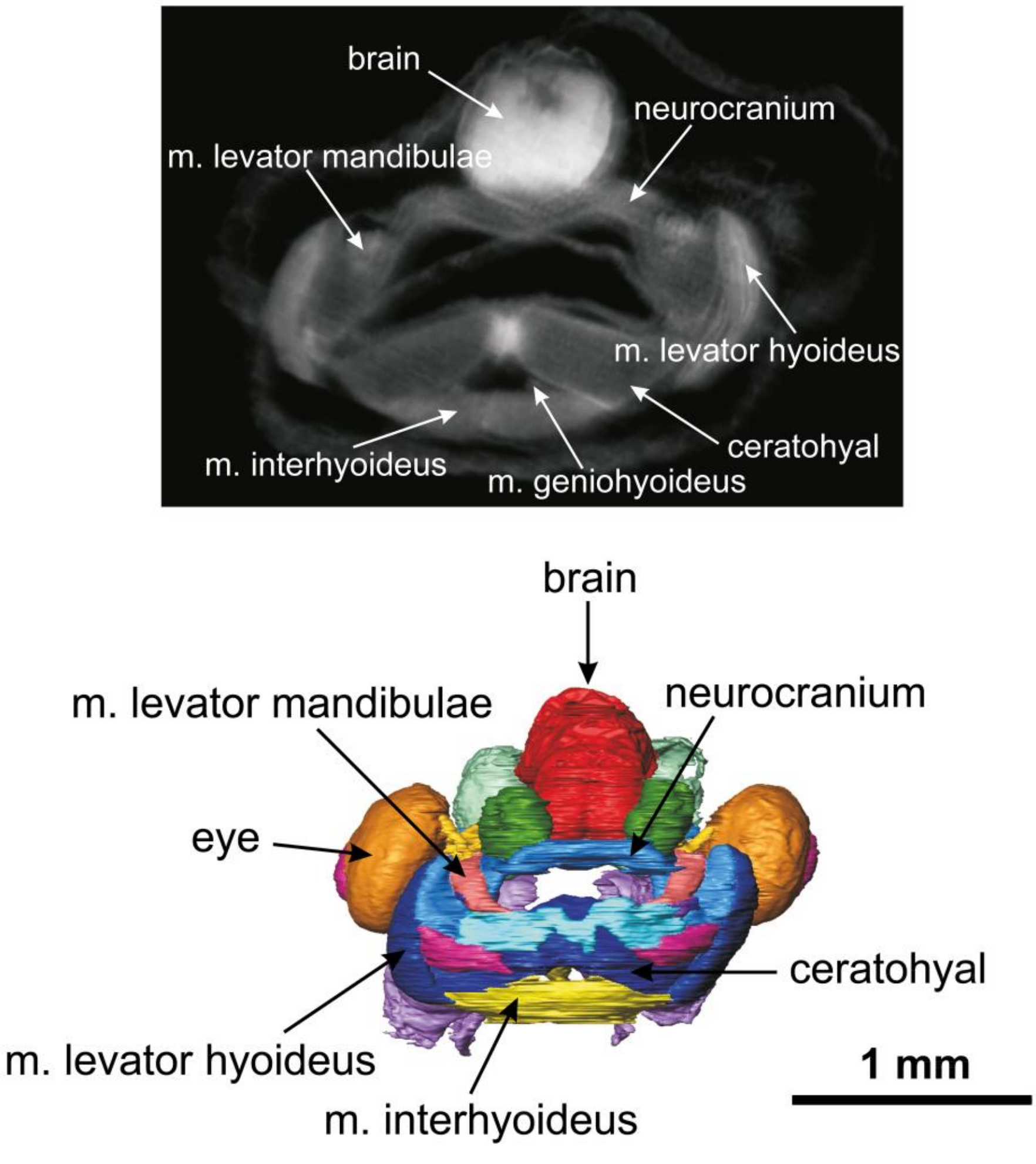

**Figure 9. (Top)** A transverse OPFOS cross section through a tadpole head with indications of the different tissue types. **(Bottom)** A 3-D reconstruction of the entire functionally segmented OPFOS image data stack (sensory organs, muscles, cartilage and neuronal structures in different colors) (frontal view). Voxel size 1.5x1.5x3 µm.

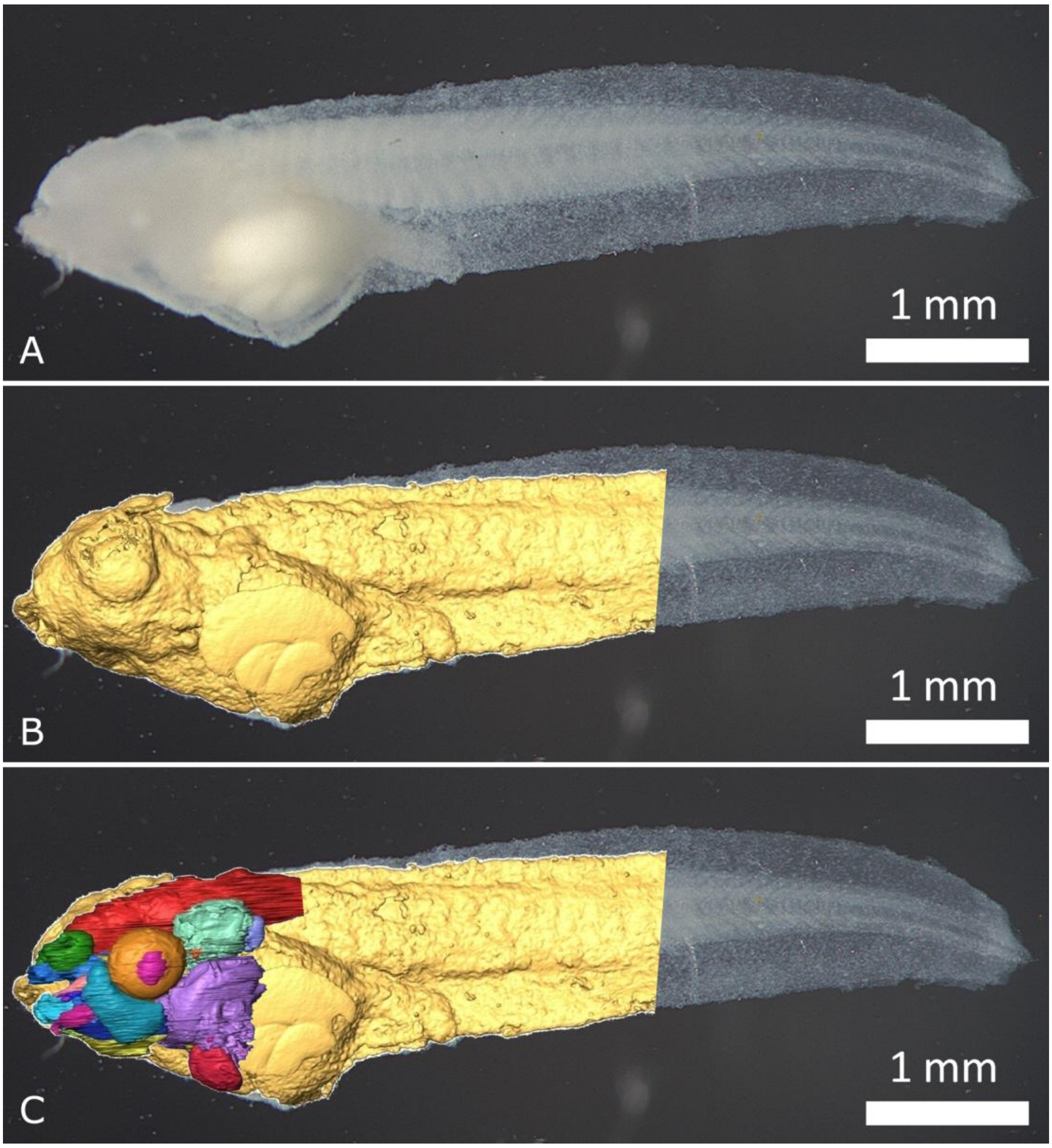

**Figure 10. (A)** A photograph of the tadpole after bleaching. **(B)** The photograph is superposed with the OPFOS surface mesh of the tadpole head and body. **(C)** Color coded functional segmentation of individual organs, cf. Figure 9.

## LSFM drawbacks

The elaborate specimen preparation required in OPFOS and other LSFM techniques is a major disadvantage. The method is considered non-destructive, however, dehydration removed all water content and decalcification did the same with calcium. It is clear that shrinkage is thus unavoidable and in the same order of magnitude as serial histological sectioning [16,18,19].

The accuracy of measurements based on OPFOS sections depends greatly on the quality of the transparency of the sample, and thus on the bleaching, dehydration and decalcification process. Dark or dense regions in the sample, remaining water content or calcium atoms refract or scatter laser light, leading to out-of-focus illumination and blurring. Furthermore, the illuminating light sheet entering the sample from one side can be partially absorbed in dense regions resulting in loss of excitation light and fluorescence on the far side of the region. Remaining pigment or zones of less(er) transparency also create this kind of shadows. These stripes or shadow line artifacts are a typical drawback of OPFOS-like techniques. Solutions for these stripes have been implemented, cf. the following section.

Finally, it is important to keep the distance and the amount of refractive material constant between the laser light sectioning plane and the observation lens when sectioning different depths. By translating the refraction-index-matched sample within the Spalteholz-filled specimen chamber orthogonally to the light sheet [5,7], or by rotating it within the chamber [1,12], this condition is fulfilled. However, when the entire specimen holder is moved to scan an image stack, the focus will degrade as the focal plane and sectioning plane no longer match [8].

## The OPFOS family

Optical sectioning with a plane of light was initiated in 1903 by Siedentopf and Zsigmondy [3]. Their Ultramicroscopy light sheet idea was revived 90 years later by Voie et al. with the OPFOS microscope [1,12,20]. This invention initiated the LSFM field but awareness and growth of the field only followed after the 2004 publication of the Single or Selective Plane Illumination Microscope (SPIM) in *Science* by Huisken et al. [5]. Before in 2002, Fuchs et al. also built an LSFM device but not for tissue sectioning microscopy [21]. Their Thin Laser Light Sheet Microscope (TLSM) was used for identification of aquatic microbes in oceanic seawater (rather in the manner of Zsigmondy's Ultramicroscope for colloidal gold particles [22]). The SPIM implementation was developed at Stelzer's EMBL lab in Heidelberg (Germany) and quickly led to many new and improved designs. The SPIM authors claim to have invented light sheet illumination and orthogonal observation independently from Ultramicroscopy and OPFOS – though being aware of and citing OPFOS in 1995 [23] – based on their work on oblique confocal (theta) microscopy [24]. SPIM omits the Spalteholz clearing method which allows to use living animal embryos that possess a natural degree of transparency, like Medaka (*Oryzias latipes*) and fruit fly (*Drosophila melanogaster*) embryos embedded in agarose. Sometimes multiple SPIM image stacks are recorded between which the sample was rotated, and post-processing combines them into one high-quality multiview reconstruction.

In 2007, Dodt et al. published a new LSFM setup in *Nature Methods*, again called Ultramicroscopy in honor of Zsigmondy, countering the inherent problem of stripe artifacts. The authors added optical components to illuminate the sample simultaneously from opposing sides, effectively reducing the presence of stripes in the images. The Dodt group focuses on visualizing brain tissue. The same year Huisken et al. also started implementing bi-directional sheet illumination (and two constantly pivoting cylindrical lenses) to reduce these stripes, but his multidirectional SPIM or mSPIM setup measures each light sheet consecutively [6]. The resulting two image datasets are computationally combined yielding an image with minimal stripes. Another innovation in mSPIM is related to the quality of light sheet illumination. Each mSPIM cylindrical lens focuses laser light to a horizontal line into the back focal plane of microscope objective lens. Hence, the quality and aberrations of the light sheet is determined by the well-corrected objective and not by the cylindrical lens.

Whenever using cylindrical lenses for light sheet generation, the resulting parabolic focus can only be approximated as a plane over a length described by the confocal parameter, cf. the section on OPFOS setup. The minimal beam waist thickness of the parabolic focus widens near the edges of the confocal parameter with a factor $\sqrt{2}$ . Consequently the light plane has no constant thickness and thus no constant sectioning resolution. Furthermore, a trade-off exists between the length of the confocal parameter and the thickness of the plane. Large(r) macroscopic samples require a large confocal parameter and consequently a thick sectioning plane and low sectioning resolution. Buytaert and Dirckx resolved this problem in 2007 by line-scanning the sample across the minimal beam waist, and stitching the section image columns together [7]. In this way, the confocal parameter is allowed to be small, producing a thin sectioning plane and high sectioning resolution. Their implementation was called High-Resolution OPFOS or HR-OPFOS. The newest version of HR-OPFOS incorporates bi-directional sheet illumination as in Ultramicroscopy, cf. Figure 2 [4].

In 2008, three new LSFM versions were developed. Holekamp et al. fixed the light sheet illumination unit to the observation objective [25]. This implementation was referred to as Objective-Coupled Planar Illumination (OCPI) used for living brain imaging. Dunsby et al. used a one high numeric aperture lens in his Oblique Plane Microscope (OPM) to both illuminate the sample with an oblique light sheet and observe the fluorescence [26]. Finally, Keller et al. introduced a new method to generate a light sheet. A tilting mirror rapidly scans a micrometer thin *spherical* focus of laser light into a plane [27]. The method is called Digital Scanned Laser Light Sheet Fluorescence Microscopy (DSLM).

Thin-Sheet Laser Imaging Microscopy (TSLIM) by Santi et al. incorporates many improved features of the previous devices [28], namely the bi-directional light-sheet illumination from Ultramicroscopy; the image stitching idea from HR-OPFOS; and the combination of cylindrical lenses with aberration corrected objectives from mSPIM.

Finally Mertz and Kim developed the HiLo LSFM system [29]. This DSLM-based device counters sample-induced scattering and aberrations that broaden the thickness of the sheet illumination. Through sequential uniform and structured sheet illumination, out-of-focus background can be identified and rejected in post-processing, improving the image quality.

## Commercial devices

The long lasting lack of a commercial LSFM device is responsible for the many different implementations of the basic method, and for the unfamiliarity of researchers with the technique in certain fields [11]. This is all about to change since now LSFM microscopes have become commercially available.

*Carl Zeiss* showed a prototype of a commercial LSFM device, named SPIM, at the First LSFM meeting in 2009 in Dresden (Germany). *Zeiss* is still preparing the launch of their system, but *LaVision BioTec* already launched the Ultramicroscope (in collaboration with Dodt) near the end of 2009 at Neuroscience in Chicago (US). The samples are limited to less than one cubic centimeter and require clearing. The device is optimized to image juvenile mouse brains, and complete mouse and fruit fly embryos. *LaVision* acknowledges the initial Ultramicroscopy idea by Zsigmondy, and OPFOS by Voie as being the first tissue microscopy implementation.

## Conclusions

We have shown with several applications that the OPFOS (and derived) methods, better known as Light Sheet based Fluorescence Microscopy or LSFM, are a valuable addition for sectional imaging and three-dimensional modeling of anatomic structures. LSFM has the major advantage that the virtual slices are automatically and perfectly aligned, making it easy to generate 3-D models from them. Microscopy techniques are either focusing on flexibility, imaging depth, speed or resolution. LSFM has all these benefits according to device manufacturers and the LSFM scientific community. Specimens containing both bone and soft tissue and ranging from microscopic till small macroscopic in size, can be studied with LSFM, with application in biomedical and life sciences. This microscopy method is relatively new, conceptually simple but powerful. Researchers can easily build their own setup, and even the first commercial devices are becoming available.

## Acknowledgments

We thank A. Voie, P. Santi and U. Schröer for sharing information on OPFOS, the LSFM field and the commercial Ultramicroscope device. We gratefully acknowledge the financial support of the Research Foundation – Flanders (FWO), the Fondation belge de la Vocation, the University of Antwerp, and the GOA project (01G01908) of Ghent University.

# References

[1] A.H. Voie, D.H. Burns, and F.A. Spelman, “Orthogonal-plane fluorescence optical sectioning: three-dimensional imaging of macroscopic biological specimens.,” *Journal of microscopy*, vol. 170, 1993, pp. 229-236.

[2] W. Spalteholz, *Über das Durchsichtigmachen von menschlichen und tierischen Präparaten*, Verlag Hirzel, 1911.

[3] H. Siedentopf and R.A. Zsigmondy, “Uber Sichtbarmachung und Größenbestimmung ultramikoskopischer Teilchen, mit besonderer Anwendung auf Goldrubingläser,” *Annalen der Physik*, vol. 315, 1903, pp. 1-39.

[4] J.A.N. Buytaert and J.J.J. Dirckx, “Tomographic imaging of macroscopic biomedical objects in high resolution and three dimensions using orthogonal-plane fluorescence optical sectioning.,” *Applied optics*, vol. 48, Feb. 2009, pp. 941-918.

[5] J. Huisken, J. Swoger, F.D. Bene, J. Wittbrodt, and E.H.K. Stelzer, “Live Embryos by Selective Plane Illumination Microscopy,” *Science*, vol. 305, 2004, pp. 1007-1009.

[6] J. Huisken and D.Y.R. Stainier, “Even fluorescence excitation by multidirectional selective plane illumination microscopy (mSPIM).,” *Optics letters*, vol. 32, Sep. 2007, pp. 2608-2610.

[7] J.A.N. Buytaert and J.J.J. Dirckx, “Design and quantitative resolution measurements of an optical virtual sectioning three-dimensional imaging technique for biomedical specimens, featuring two-micrometer slicing resolution.,” *Journal of biomedical optics*, vol. 12, 2007, p. 014039.

[8] H.-U. Dodt, U. Leischner, A. Schierloh, N. Jahrling, C.P. Mauch, K. Deininger, J.M. Deussing, M. Eder, W. Zieglgansberger, and K. Becker, “Ultramicroscopy: three-dimensional visualization of neuronal networks in the whole mouse brain,” *Nature methods*, vol. 4, 2007, pp. 331-336.

[9] R. Hofman, J.M. Segenhout, J.A.N. Buytaert, J.J.J. Dirckx, and H.P. Wit, “Morphology and function of Bast’ s valve: additional insight in its functioning using 3D-reconstruction.,” *European archives of oto-rhino-laryngology*, vol. 265, Feb. 2008, pp. 153-157.

[10] J. Buytaert, E. Descamps, D. Adriaens, and J. Dirckx, “Orthogonal-Plane Fluorescence Optical Sectioning : a technique for 3-D imaging of biomedical specimens,” *Microscopy: Science, Technology, Applications and Education*, A. Méndez-Vilas and J. Díaz, eds., FORMATEX, 2010, pp. 1356-1365.

[11] P.A. Santi, “Light sheet fluorescence microscopy: a review,” *The journal of histochemistry and cytochemistry : official journal of the Histochemistry Society*, vol. 59, Feb. 2011, pp. 129-138.

[12] A.H. Voie, “Imaging the intact guinea pig tympanic bulla by orthogonal-plane fluorescence optical sectioning microscopy.,” *Hearing research*, vol. 171, Sep. 2002, pp. 119-128.

[13] S.P. Tinling, R.T. Giberson, and R.S. Kullar, “Microwave exposure increases bone demineralization rate independent of temperature.,” *Journal of microscopy*, vol. 215, Sep. 2004, pp. 230-235.

[14] C.F.A. Culling, “Spalteholz Technique,” *Handbook of Histopathological and Histochemical Techniques*, Butterworths, 1974, pp. 550-552.

[15] V. Ermolayev, M. Friedrich, R. Nozadze, T. Cathomen, M. a Klein, G.S. Harms, and E. Flechsig, “Ultramicroscopy reveals axonal transport impairments in cortical motor neurons at prion disease.,” *Biophysical journal*, vol. 96, Apr. 2009, pp. 3390-8.

[16] J.A.N. Buytaert, W.H.M. Salih, M. Dierick, P. Jacobs, and J.J.J. Dirckx, “Realistic 3-D computer model of the gerbil middle ear, featuring accurate morphology of bone and soft tissue structures,” *JARO-Journal of the Association for Research in Otolaryngology*, vol. submitted, 2011.

[17] S. Van Wassenbergh, G. Roos, A. Genbrugge, H. Leysen, P. Aerts, D. Adriaens, and A. Herrel, “Suction is kid’ s play: extremely fast suction in newborn seahorses.,” *Biology letters*, vol. 5, Apr. 2009, pp. 200-203.

[18] R. Hofman, J.M. Segenhout, and H.P. Wit, “Three-dimensional reconstruction of the guinea pig inner ear, comparison of OPFOS and light microscopy, applications of 3D reconstruction,” *Journal of microscopy*, vol. 233, Feb. 2009, pp. 251-257.

[19] J. Lane and Z. Ráliš, “Changes in dimensions of large cancellous bone specimens during histological preparation as measured on slabs from human femoral heads,” *Calcified Tissue International*, vol. 35, 1983, p. 1–4.

[20] A.H. Voie and F.A. Spelman, “Three-dimensional reconstruction of the cochlea from two-dimensional images of optical sections,” *Computerized Medical Imaging and Graphics*, vol. 19, 1995, p. 377–384.

[21] E. Fuchs, J. Jaffe, R. Long, and F. Azam, “Thin laser light sheet microscope for microbial oceanography.,” *Optics express*, vol. 10, Jan. 2002, pp. 145-54.

[22] R.A. Zsigmondy, “Properties of colloids,” *Nobel lecture in chemistry*, 1926, pp. 1-13.

[23] E.H.K. Stelzer, S. Lindek, S. Albrecht, R. Pick, G. Ritter, N.J. Salmon, and R. Stricker, “A new tool for the observation of embryos and other large specimens: confocal theta fluorescence microscopy,” *Journal of microscopy*, vol. 179, 1995, p. 1–10.

[24] S. Lindek and E.H.K. Stelzer, “Confocal theta microscopy and 4Pi-confocal theta microscopy,” *Proceedings of SPIE, Vol.2184: Three-Dimensional Microscopy*, SPIE International Society for Optical Engineering, 1994, p. 188–188.

[25] T.F. Holekamp, D. Turaga, and T.E. Holy, "Fast three-dimensional fluorescence imaging of activity in neural populations by objective-coupled planar illumination microscopy," *Neuron*, vol. 57, Mar. 2008, p. 661–672.

[26] C. Dunsby, "Optically sectioned imaging by oblique plane microscopy.," *Optics express*, vol. 16, Dec. 2008, pp. 20306-16.

[27] P.J. Keller, A.D. Schmidt, J. Wittbrodt, and E.H.K. Stelzer, "Reconstruction of zebrafish early embryonic development by scanned light sheet microscopy.," *Science (New York, N.Y.)*, vol. 322, Nov. 2008, pp. 1065-9.

[28] P.A. Santi, S.B. Johnson, M. Hillenbrand, P.Z. GrandPre, T.J. Glass, and J.R. Leger, "Thin-sheet laser imaging microscopy for optical sectioning of thick tissues," *BioTechniques*, vol. 46, Apr. 2009, pp. 287-94.

[29] J. Mertz and J. Kim, "Scanning light-sheet microscopy in the whole mouse brain with HiLo background rejection," *Journal of biomedical optics*, vol. 15, 2011, p. 016027.